\documentclass[sigconf]{acmart}
\acmConference[ICSE 2024]{46th International Conference on Software Engineering}{April 2024}{Lisbon, Portugal}

\usepackage{lipsum}

\AtBeginDocument{%
  \providecommand\BibTeX{{%
    \normalfont B\kern-0.5em{\scshape i\kern-0.25em b}\kern-0.8em\TeX}}}

\copyrightyear{2024}
\acmYear{2024}
\setcopyright{rightsretained}
\acmConference[ICSE '24]{2024 IEEE/ACM 46th International Conference on
Software Engineering}{April 14--20, 2024}{Lisbon, Portugal}
\acmBooktitle{2024 IEEE/ACM 46th International Conference on Software
Engineering (ICSE '24), April 14--20, 2024, Lisbon,
Portugal}\acmDOI{10.1145/3597503.3639183}
\acmISBN{979-8-4007-0217-4/24/04}

\usepackage{todonotes}

\usepackage{amssymb}
\usepackage{amsmath, amsfonts}

\usepackage{graphicx}
\usepackage{textcomp}
\usepackage{xcolor}
\usepackage{hyperref}
\usepackage{cleveref}
\usepackage{xspace}
\usepackage{paralist}
\usepackage{multirow}
\usepackage{subcaption}
\usepackage{tikz}
\usepackage{courier}
\usepackage{listings} 
\usepackage{array}
\usepackage{graphics}
\usepackage{lipsum}

\usepackage{etoolbox}
\apptocmd{\sloppy}{\hbadness 10000\relax}{}{}

\newcommand{\etal}{\emph{et al.}\xspace}
\newcommand{\etc}{\emph{etc.}\xspace}
\newcommand{\ie}{\emph{i.e.},\xspace}
\newcommand{\eg}{\emph{e.g.},\xspace}
\newcommand{\viz}{\emph{viz.}\xspace}

\newcommand{\asap}{${\mathcal ASAP}$\xspace}

\definecolor{Gray}{gray}{0.3}
\tikzstyle{mybox} = [draw=black, very thick, rectangle, rounded corners, inner ysep=5pt, inner xsep=5pt, fill=gray!20]

\begin{document}

\title{Automatic Semantic Augmentation of Language Model Prompts \\(for Code Summarization)}

\author{Toufique Ahmed}
\affiliation{%
  \institution{University of California, Davis}
  \city{Davis}
  \state{California}
  \country{USA}
  \postcode{95616}}
\email{tfahmed@ucdavis.edu}

\author{Kunal Suresh Pai}
\affiliation{%
  \institution{University of California, Davis}
  \city{Davis}
  \state{California}
  \country{USA}
  \postcode{95616}}
\email{kunpai@ucdavis.edu}

\author{Premkumar Devanbu}
\affiliation{%
  \institution{University of California, Davis}
  \city{Davis}
  \state{California}
  \country{USA}
  \postcode{95616}}
\email{ptdevanbu@ucdavis.edu}

\author{Earl T. Barr}
\affiliation{%
  \institution{University College London \& Google Brain}
  \city{London}
  \country{UK}
  \postcode{95616}}
\email{e.barr@ucl.ac.uk}




\begin{abstract}

Large Language Models (LLM) are a new class of computation engines, ``programmed'' via prompt engineering. Researchers are still learning how to best ``program'' these LLMs to help developers.  
We start with the intuition that developers
tend to consciously and unconsciously collect
semantics facts, from the code, while working. 
Mostly these are shallow, simple facts arising 
from a quick read.  For a function, such facts might include parameter and local variable names, return expressions, simple pre- and post-conditions, and basic control and data flow, \etc

One might assume that the powerful multi-layer
architecture of transformer-style LLMs makes them \emph{implicitly} capable of doing this simple level of
``code analysis'' and extracting such information, while processing code: but are they, really? If they aren't, could \emph{explicitly} adding this information help? 
Our goal here is to investigate this question, using the \emph{code summarization task} and
evaluate whether automatically augmenting an LLM's prompt with semantic facts \emph{explicitly}, actually helps.

Prior work shows that LLM performance on code summarization benefits from embedding a few
code \& summary exemplars in the prompt, before the code to be summarized. 
While summarization performance has steadily progressed since the early days, there is still room for improvement: LLM performance on code summarization  still lags its performance on natural-language tasks like translation and text summarization.

We find that adding semantic facts to the code in the prompt
 actually does help! This approach improves performance in several different
settings suggested by prior work, including for \emph{three} different Large Language Models. In most cases, we see improvements, as measured by a range of commonly-used metrics; 
for the PHP language in the challenging CodeSearchNet dataset, this augmentation
actually yields performance surpassing 30 BLEU\footnote{Scores of 30-40 BLEU are considered ``Good" to "Understandable" for natural language translation; see \url{https://cloud.google.com/translate/automl/docs/evaluate}.}. In addition, we have also found that including semantic facts yields a substantial enhancement in LLMs' line completion performance. 


\end{abstract} 




\maketitle

\section{Introduction}
Large language models (LLMs) often outperform smaller, custom-trained models on tasks, especially when prompted with a "few-shot" set of exemplars. LLMs are pre-trained on a self-supervised (masking or de-noising) task, using vast amounts of data, and exhibit surprising emergent behaviour as training data and parameter counts are scaled up.  They excel at many  tasks with few-shot (or even zero-shot) learning: with just
a few exemplar input-output pairs inserted first in the prompt, the models can generate very good outputs for a given input! Few-shot learning works so well with LLMs that it is unclear whether sufficient task-specific data can ever be gathered to train a customized model to rival their performance~\cite{brown2020language,ahmed2022few}. LLMs are ushering in a new era, where prompt engineering, to carefully condition the input to an LLM to tailor its massive, but generic capacity, to specific tasks, will become a new style of programming, placing new demands on software engineers. 

We propose \emph{Automatic Semantic Augmentation of Prompts} (\asap),  a new method for constructing prompts for software engineering tasks.  The \asap method rests on an analogy:  an effective prompt for an LLM, for a task, relates to the facts a developer thinks about when manually performing that task. In other words, we hypothesize that prompting an LLM 
with the syntactic and semantic facts a developer considers when manually performing a task \emph{will} improve LLM performance on that task. 
To realise this hypothesis, \asap augments prompts with semantic facts automatically extracted from the source code using \emph{semantic code analysis}. 

We illustrate the \asap methodology first on code summarization.  This task takes code, usually a function, and summarizes it using natural language; such summaries can support code understanding to facilitate requirements traceability and maintenance.
\asap uses a few-shot prompting because its effectiveness.
\asap finds relevant shots using \emph{BM25}, the current state of the art in finding few-shot exemplars that are ``semantically close'' to the target function~\cite{nashid2023retrieval}, in our case, the function-to-summarize, by querying the LLM's training data. 
When instantiating \asap for the summarization task, we equipped it to extract the following semantic facts:  the repository name, the fully qualified name of the name of the target function, its signature, the AST tags of its identifiers, and its data flow graph (\Cref{sec:aspa}). 
These facts are presented to the LLM as separate, labelled,  
fields\footnote{A full example is rather long, and is
included in the repository due to paper length limitations.}. 
The model is then provided with the function-to-summarize, exemplars (along with facts extracted from each), and asked
to emit a summary.  
We  confirm our hypothesis that augmenting prompts with semantic facts can improve LLM performance on the code completion task.
We evaluated \asap's benefits on the high-quality (carefully de-duplicated, multi-project) CodeSearchNet~\cite{husain2019codesearchnet} dataset. 

In summary, we find that in \emph{all cases}, our approach of automatic semantic augmentation \emph{improves average performance} on \emph{several 
commonly-used metrics}. For almost all languages,the average improvement comfortably surpasses the 2-BLEU threshold noted by Roy \emph{et al.}~\cite{roy2021reassessing}, below which BLEU results are unreliable predictors of human preference. For Go, gains are still significant, and just slightly less than 2; for PHP, we see an improvement of \emph{4.6 BLEU}, reaching a SOTA high-point of 32.73\, on the well-curated, de-duplicated, CodeSearchNet dataset. 

Our principal contributions follow:

\begin{itemize}
    \item The \asap 
    approach for software engineering tasks using facts derived from code.  
    \item We evaluate \asap on the code summarization task on the code-davinci-002, 
    text-davinci-003. 
    and GPT-3.5-turbo models against a few-shot prompting baseline built using vanilla \emph{BM25} (\Cref{fixed_few_shot}). 
    \item We find that the \asap approach statistically significantly improves LLM performance on the code summarization task.  In almost all cases, we observe statistically significant improvements of almost, or in excess of, 2 BLEU;  
    and, for PHP, we break 30 BLEU for the first time (to our knowledge) on this challenging dataset. 
    \item We find that \asap also leads to improved performance on the code-completion task. 
    
\end{itemize}

All the data, evaluation scripts, and code needed to reproduce this work will be 
available at \url{https://doi.org/10.5281/zenodo.7779196}, and can be reproduced
on any available language models. Our experiments suggest that 
\asap  works well with any language model powerful enough to leverage few-shot
prompting. 
\section{Background \& Motivation}

Large Language Models (LLM) are a transformative technology: they are essentially a new kind of computation engine, requiring a new form of programming, called prompt engineering.
We first contextualise \asap, our contribution to prompt engineering.  Finally, we discuss code summarization as a sample problem to  demonstrate \asap's effectiveness. 

\subsection{Few-shot Learning in Software Engineering} 

LLMs are now widely used in Software Engineering for
many different problems: code generation~\cite{chen2021evaluating,jain2022jigsaw},
testing~\cite{kang2023large,lemieux2023codamosa}, mutation generation~\cite{bareiss2022code}, program repair~\cite{jiang2023impact,nashid2023retrieval,joshi2022repair,fan2022automated},
incident management~\cite{ahmed2023recommending},
and even code summarization~\cite{ahmed2022few}.
Clearly, tools built on top of pre-trained LLM are
advancing the state of the art.
Beyond their raw performance at many tasks, two key factors govern the growing dominance of pretrained LLM,
both centered on cost.
First, training one's own large model, or even extensively fine-tuning a pre-trained LLM, requires expensive hardware.  Second, generating a supervised dataset for many important software engineering tasks is difficult and time-consuming, often beyond the sources of
all but the largest organizations. 

In contrast to overall LLM trends, there are some smaller models, specialized for code, that have gained popularity, \eg Polycoder~\cite{xu2022systematic} or Codegen~\cite{nijkamp2022codegen}.
Despite these counterpoints, we focus on LLM rather than small models, because, 
while small models can be fine-tuned, they 
don't do very well at few-shotting, and thus are not helpful when only
small amounts of data are available. 
The few-shot approach is key because it brings into reach 
many problems, like code summarization, for which 
collecting sufficient, high-quality, project- or domain-specific training data 
to train even small models from scratch is challenging.

With few-shot learning, 
the actual model parameters remain unchanged. 
Instead, we present a few problem instances along with solutions (\emph{i.e.,} problem-solution pairs
as ``the exemplars")
to a model and ask it to complete 
the answer for the last instance ("the test input"), for which we do not provide a solution. Thus 
with each $exemplar$ consisting of an $\langle \mathit{input}, \mathit{output} \rangle$ pair, and just a 
test-input $\mathit{input}_t$ (without the corresponding, desired $\mathit{output}_t$), the final prompt looks like: 
\[\mathit{prompt} \leftarrow \mathit{exemplar}_1 \mid\mid \mathit{exemplar}_2 \mid\mid \mathit{exemplar}_3 \mid\mid input_t \]
With this prompt, the LLM generates $output_t$, mimicking the input-output behavior 
illustrated by the exemplars in the prompt. In practice, this approach performs quite well. 

When it works, few-shotting allows us to automate even purely manual problems,
since generating a few exemplar samples is relatively easy.
In this paper, we experiment with the 
code-davinci-002 model.  We discuss models in more detail in \autoref{model}.
 
\subsection{Prompting LLMs to Reason}
Human Reasoning involves using evidence, logical thinking, and arguments to make judgments or arrive at conclusions~\cite{qiao2022reasoning,huang2022towards}.
Natural language processing (NLP) researchers have developed approaches to 
reason about specific scenarios and improve performance. Approaches like "Chain of thought"~\cite{wei2022chain} and "step-by-step"~\cite{kojima2022large} require generating intermediate results (``lemmas") and utilizing them in the task at hand. Such approaches appear to work on simpler problems like school math problems even without providing them with ``lemmas" , because, for these problems, models are powerful enough to generate their own ``lemmas"; in some cases just adding ``let's think step by step" seems sufficient (Kojima \emph{\etal}~\cite{kojima2022large}). 

We tried an enhanced version of the ``step-by-step'' prompt,  with few-shots, on code summarization. We found that the model  under-performed (getting about 20.25 BLEU), lower even than our vanilla BM25 baseline (24.97 BLEU). With zero-shot Kojima-style ``step by step'' prompt, the models perform even worse. To induce the model to generate steps, and finally a summary, we framed the problem as chain of thought, \emph{and} included few-shot samples containing both intermediate steps (``lemmas") and final comments. 
The reasoning is that, on the (usually challenging) code-related tasks, models need to explicitly be given intermediate ``lemmas", \emph{derived from code}, to be able to reason effectively about most software engineering tasks, which tend to be more complex and varied than school maths.

Fortunately, 
mature tools for code analysis are available. We can readily derive 
``lemmas", \emph{viz.,} analysis products, using code analysis tools, rather than 
expecting the models to (perhaps implicitly) derive them, during on-task performance. We directly
embed analysis products into the prompt we give the language model, and evaluate the benefits of such analysis products. 
The information we derive and add are based on our own intuitions about the kinds of ``lemmas" that  developers consciously or unconsciously consider as they seek to understand and summarize code. 
 
We find that providing such information improves  
LLM performance. 
We remind the reader that most work involving large language models (LLMs) usually uses some form of prompt engineering to boost performance. In this paper, we show that the \asap approach, which augments prompts 
with code analysis products, improves on previous prompting approaches. 

\subsection{Summarizing Code}
Well-documented code is much easier to maintain; thus, experienced developers 
usually add, \emph{e.g.}, function summary headers. 
However, summary comments may become outdated, as projects evolve~\cite{briand2003software,forward2002relevance}. 
Automated code summarization is thus a well-motivated task, which has attracted a
great deal of attention; and considerable progress (albeit incremental, over many years) has been made. 
Initially, template-based approaches were popular~\cite{eddy2013evaluating,haiduc2010supporting,haiduc2010use,sridhara2010towards,rodeghero2014improving}; 
however, creating a list of templates with good coverage is very challenging. 
Later, researchers focused on the retrieval-based (IR) approach~\cite{eddy2013evaluating,haiduc2010supporting,haiduc2010use,rodeghero2014improving}, where existing code (with a summary) is retrieved based on similarity-based metrics. 
However, this promising approach only worked if a similar code-comment pair could be found 
in the available pool. 

Meanwhile, the similarity of code summarization to Neural Machine Translation (NMT), 
(one can think of generating an English summary of code as producing
a representation of ``the same meaning in a different language'')
led to research that adapted Neural Machine Translation (NMT) to code summarization. 
Numerous studies have been conducted in this area~\cite{ahmad2020transformer, leclair2019neural,iyer2016summarizing,hu2018summarizing}.
Some have combined previous approaches, such as template-based and retrieval-based approaches, using neural models~\cite{zhang2020retrieval}, and have reported promising results. Such neural methods for NLP have vastly improved, due to the 
Transformer architectural style. 

Until recently, pre-trained language models such as CodeBERT, CodeT5, and CodeT5+ performed best for code summarization. 
However,  Large Language Models (LLMs) now typically outperform smaller pre-trained models on many problems.  
Ahmed \& Devanbu~\cite{ahmed2022few} report that LLMs can outperform pre-trained language models with a simple prompt consisting of just a few  samples already in the same project; this work illustrates
the promise of careful 
construction of prompt structures (\emph{c.f.} ``prompt engineering'').  
We present \asap here as another general principle of prompt engineering. 
We emphasize, again, that
progress in code summarization (and other applications of AI to SE, such as code patching, defect
detection, testing \emph{etc}) has been incremental, as in the field of NMT, where practical, usable translation systems took decades
to emerge. Thus \emph{incremental advances are still needed, and helpful,}
and we contribute our work to this long-term enterprise. 

\section{Dataset \& Methodology}
\label{methodology}
We now discuss our dataset, models, and methodology.

\begin{figure}[!t]
    \centering
    \includegraphics[scale=0.38]{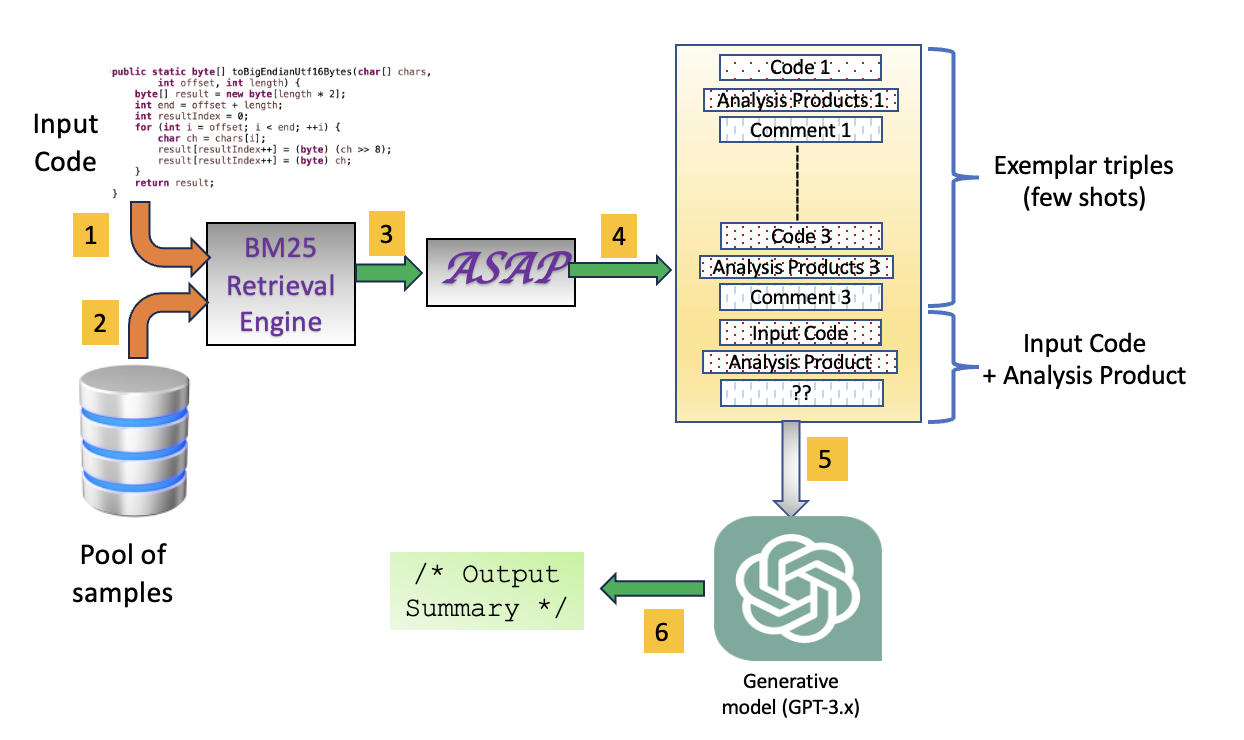}
    \vspace{-0.1in}
    \caption{Different steps of \asap. (1) Input code and (2) Pool of samples are given the BM25 engine, which matches the given input code against the pool and (3) retrieves best-matching samples, \viz 3 input+output pairs. These examples
    are processed by \asap to produce a prompt (4) including 3 exemplars.  
    Each exemplar includes a function definition, the results of analyzing that definition, and its associated comment; the input code is finally appended, along  with its analysis product. Exemplar details are in Figure \ref{component}. The final prompt 
    is sent via API call (5) to the GPT-3.x model; the returned output, \eg summary (6) 
    is returned by GPT-3x.}
    \label{pipeline}
    \vspace{-0.2in}
\end{figure}

\begin{figure}[!t]
    \centering
    \includegraphics[scale=0.18]{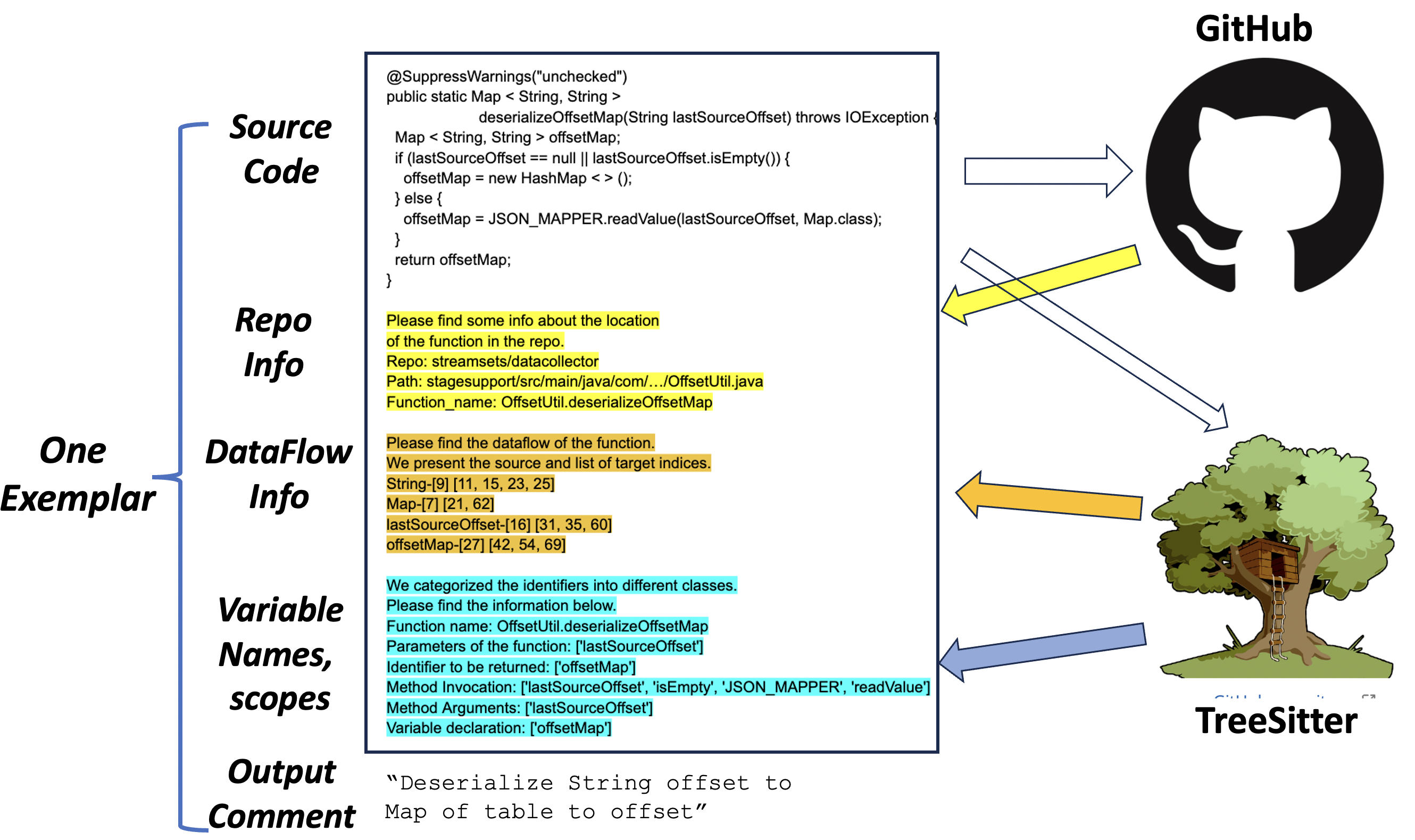}
    \caption{Components of an \asap Exemplar. Source Code and Output Comment are extracted from the retrieved pool sample. The Repo info is derived from the source code using GitHub; the Dataflow Info and tagged
    Identifiers with labels is obtained from an analysis using Treesitter.}
   \label{component}
       \vspace{-0.2in}
\end{figure}

\subsection{Dataset}

Our experiments  use the widely used CodeSearchNet~\cite{husain2019codesearchnet} dataset;  CodeSearchNet was constructed by extracting the first paragraph of the function prefix documentation, subject to some restrictions (\emph{e.g.} length). 
It is a carefully de-duplicated, multi-project dataset, which allows (more demanding)
cross-project testing. 
De-duplication is key:  Code duplication in machine learning models can deceptively inflate performance metrics a lot, 
when compared to de-duplicated datasets~\cite{allamanis2019adverse,lopes2017dejavu,shi2022evaluation}.

It is part of the CodeXGLUE~\cite{lu2021codexglue} benchmark, which comprises
14 datasets for 10 software engineering tasks. Many models have been evaluated on this dataset. 
CodeSearchNet contains thousands of samples 
from six different programming languages 
(\ie Java, Python, JavaScript, Ruby, Go, PHP).
However, we did not use the entire test dataset, which would have been prohibitively expensive and slow
using ours models API endpoints; instead, we selected 
1000 samples\footnote{Please see experimental power discussion in \Cref{sec:sample-size}.} 
uniformly at random from each language. Since the original dataset is cross-project and we sampled it uniformly, our subsample includes cross-project data. 
In addition, we subsetted this dataset for same-project few-shotting, following Ahmed and Devanbu~\cite{ahmed2022few}:  
we sort same-project data by creation date (using {\small\tt git blame}). Now, we use
the temporal order to make sure that 
only temporally \emph{earlier} samples are used the few-shot exemplars; this is realistic, since only older, already existing data is available for use. 
We will delve deeper into this same-project dataset in Section~\ref{same-project-summary}.

As mentioned earlier, we don't use any parameter-changing training
on the model; we just insert a few exemplars selected from the training subset into the few-shot prompt. 
\Cref{tbl:data} lists the count of training \& test samples used in our experiments.

\begin{table}[t]
\centering
\resizebox{.65\columnwidth}{!}{%
\renewcommand{\arraystretch}{1.2}
\begin{tabular}{lrr}
\toprule
\multicolumn{1}{c}{Language} & \#of Training Samples & \#of Test Samples \\ 
\midrule
Java                         & 164,923               & 1000               \\
Python                       & 251,820               & 1000              \\
Ruby                         & 24,927                & 1000               \\
JavaScript                   & 58,025                & 1000              \\
Go                           & 167,288               & 1000              \\
PHP                          & 241,241               & 1000   \\ 
\bottomrule           
\end{tabular}
}
\vspace{0.05in}
\caption{Number of training and test samples.}
\label{tbl:data}
\vspace{-0.40in}
\end{table}

\subsection{The Models}
\label{model}

In earlier work,  transformer-based pre-trained 
language models offered significant gains, in both NLP and software engineering. 
Pre-trained language models can be divided 
into three categories: encoder-only, encoder-decoder, and decoder-only models.
While encoder-decoder models have initially 
shown success on many tasks, decoder-only LLMs are now more scaleable and effective for numerous tasks.

\paragraph{Encoder-Decoder model}
BERT is one of the earliest pre-trained language models~\cite{devlin2018bert}; it was pre-trained on two self-supervised tasks: Masked Language Modeling (MLM) and Next Sentence Prediction (NSP).
Later, RoBERTa~\cite{liu2019roberta} was introduced with some minor modifications to BERT. Using only MLM training, it outperforms BERT.
CodeBERT~\cite{feng2020codebert} and GraphCodeBERT~\cite{guo2020graphcodebert} introduced these ideas to Software Engineering. Although CodeBERT and GraphCodeBERT are encoder-only models, they can be applied to code summarization after fine-tuning, cascaded to a decoder trained during fine-tuning.
Ahmed \& Devanbu report that polyglot models, which are fine-tuned with multilingual data, outperform their monolingual counterparts~\cite{ahmed2022multilingual}. 
They also report that identifiers play a critical role in code summarization tasks. PLBART~\cite{ahmad2021unified}, CodeT5~\cite{wang2021codet5}, and CodeT5+~\cite{wang2023codet5+} also include pre-trained \emph{decoders} and are reported to work well for code summarization tasks. 
More recently, very large scale (decoder-only) auto-regressive LLMs (with 175B+ parameters) have been found to be successful at code summarization with few-shot learning, without any explicit training.
In the next section, we will briefly introduce the three OpenAI models we considered for our experiments.

\paragraph{Decoder-only model}
In generative pre-training, the task is to auto-regressively predict the next token given the previous tokens moving from earlier to later. This unidirectional auto-regressive training prevents the model from pooling information from future tokens. 
The newer generative models such as GPT~\cite{radford2018improving}, GPT-2 \cite{radford2019language} and GPT-3 \cite{brown2020language},
are also trained in this way, but they have more parameters, and are trained on much larger datasets. 
Current large language models, such as GPT-3, have around (or more than) 175B parameters. These
powerful models perform so well, with few-shot prompting, that 
interest on task-specific parameter-adjustment via fine-tuning has reduced.

Codex is a GPT-3 variant, intensively trained on code and natural language comments.
The Codex family consists of two versions: Codex-Cushman, which is  smaller, with 12B parameters, and Codex-Davinci, the largest, with 175B parameters. 
The Codex model is widely used, for various tasks. 
Our experiments mostly target the Code-Davinci model, particularly Code-Davinci-002, which excels at translating natural language to code~\cite{chen2021evaluating} and supports code completion as well as code insertion\footnote{https://openai.com/}. 
Some new variants, Text-Davinci-003 \& GPT-3.5-turbo, are also available; 
unlike the Codex variants, these models  understand and generate both natural language and code. 
Although optimized for chat, GPT-3.5-turbo also performs well on traditional completion tasks. Text-Davinci-003 is a completion model like Code-Davinci-002. 
We study how our prompt enhancement works using the Text-Davinci-003 \& GPT-3.5-turbo models.

\subsection{Retrieving Exemplars from Training Data}
As noted earlier, few-shot learning works quite well, 
when used with very large models. 
We 
prompt the model with a small number of $\langle \mathit{problem},\mathit{solution}\rangle$ exemplars, and ask it to solve a new problem. 
However, carefully 
selecting exemplars for few-shot learning is helpful. 
Nashid \etal discovered that retrieval-based exemplar selection is helpful 
for problems such as assertion generation and program repair~\cite{nashid2023retrieval}. Following their recommendation,  we use the \emph{BM25} IR algorithm to select
relevant few-shot exemplars from the training set. 
\emph{BM25}~\cite{robertson2009probabilistic} is a frequency-based
retrieval method which improves upon TF-IDF~\cite{ramos2003using}.
We noted a substantial improvement over
the same fixed
exemplars in few-shot learning, as detailed in Section~\ref{fixed_few_shot}. 
Nashid \emph{\etal} compare several retrieval methods, and found \emph{BM25} works best; we
therefore use it, as well.


\subsection{Automatic Semantic Prompt Augmentation}
\label{sec:aspa} 

This section presents the three semantic facts we selected to enhance \asap's prompts and the \asap pipeline (See \autoref{component}).  The choice of these facts comes from applying our central hypothesis, \viz that augmenting 
prompts with what developers think about when working on a task, to the code summarization task.
\asap is not tied to any specific semantic facts or static analysis; it can easily incorporate others, as discussed later. 

\paragraph{Repository Name \& Path}

Augmenting prompts with domain-specific information can improve 
LLM performance on various tasks. Prior work suggests
that augmenting prompts with \emph{code} from the \emph{same repository} improves 
performance in code generation tasks~\cite{shrivastava2022repository}. 
We argue that basic repository-level \emph{meta}-information, such as the 
repository name and the complete path to the 
repository, provides additional context. 
For example, repository names like 
``\lstinline+tony19/logback-android+'', 
``\lstinline+apache/parquet-mr+'', and 
``\lstinline+ngageoint/+ \lstinline+geo-package-android+'' all 
connect a function to a specific domain (\eg 
android, apache, geo-location), which can enhance  
the understanding of the target code to be summarized. 
\autoref{component} (yellow part) presents an example 
of how we enhance the prompt with repository-level 
information. Similar to the repository name, the 
path to the function can also contribute to 
the model.

\paragraph{Tagged Identifiers}
Prior work suggests that  language models find more value in identifiers, rather than code structure, when generating code summaries~\cite{ahmed2022multilingual}. However,  identifiers  play different roles in code. Local variables, function names, 
parameters, global variables \etc, play different parts in
the functioning of the method in which they occur; a developer
reading the code is certainly aware of the roles of identifier, simply
by identifying the scope and use. 
Thus, augmenting prompts with the \emph{specific roles} of  identifiers could help the model better ``understand'' the function. We use tree-sitter to  traverse the function's AST and gather identifiers, along with their roles. \autoref{component} (blue part) presents a sample example showing how we enhanced the prompt of the function with tagged identifiers. Although the model has access to the token sequence of the code, and thus also all the identifiers, them to the model in a tagged form might a) save the model
some compute effort, and b) better condition the model's output. 

\paragraph{Data Flow Graph (DFG)}
Guo \etal introduced the GraphcodeBERT model, which uses data flow graphs (DFG) instead of syntactic-level structures like abstract syntax trees (ASTs) in the pre-training stage~\cite{guo2020graphcodebert}. 
GraphcodeBERT outperformed CodeBERT~\cite{feng2020codebert} on various software engineering (SE) tasks. We incorporate this DFG information into the few-shot exemplars; we conjecture that this provides the model a better semantic understanding of each exemplar, and the target example. \Cref{component} (orange) presents a sample showing the Data Flow Graph (DFG) we used for our experiments. Each line contains an identifier with its index and the index of the identifiers to which that particular data flows. Unlike repo and tagged identifiers, the data flow graph can be very long, making it inconvenient to add the complete data flow to the prompt. In the case of long prompts, we only kept the first 30 lines of the DFG in the prompt. In addition to identifiers, the DFG also provides a better understanding of the importance of identifiers in the function.

\paragraph{Use Case \& Completion Pipeline} 
\asap has 3 components: an LLM, a pool of available exemplars (labeled input-output pairs, 
\emph{e.g.,} code with comments),  and a static analysis tool for deriving facts from code (See \Cref{pipeline,component}).

A configuration file specifies these components.
Once configured, a developer invokes \asap on a function body $C_{in}$ (\Cref{pipeline}), for which an output (\eg, code summary) is desired.
\asap uses $C_{in}$ as a BM25 query over the its sample pool 
to get a result set of exemplar candidates $\mathit{ec}_1, \mathit{ec}_2, \ldots$, where each $\mathit{ec}_i$ is a pair of the form $\langle\mathit{input}_i, \mathit{output}_i\rangle$; 
in our context, $\mathit{input}_i$ is the function definition and $\mathit{output}_i$ is the function header comment. BM25 chooses
the $input_i$s that match best with the given $C_{in}$. 
\asap then applies program analyses to both the input $C_{in}$ and the several exemplar inputs $input_i$s, yielding
analysis products $ap_{in}$ and several $ap_i$s. 

Each exemplar $e_i$ (\Cref{component}) is the triple:
$\langle \mathit{input}_i, \mathit{ap}_i, \mathit{output}_i \rangle$,
where each triple
illustrates, for the LLM, how input source code $input_i$ relates, \emph{via} the analysis product $ap_i$, to the output $output_i$. The final prompt is then ``$e_1 \mid\mid e_2 \mid\mid  e_3 \mid\mid C_{in} \mid\mid \mathit{ap}_{in}$''.
\asap queries an LLM with that prompt, and returns the completion (\eg natural language summary).

By default, \asap is configured with analyses to extract repository info, tag identifiers, construct DFGs.  These analyses are independent and are their outputs are separately labeled in the prompt. 
For example, Figure~\ref{component} shows the output of the DFG analysis in \asap's constructed prompt.
These few shot examples, are augmented and inserted into the prompt: the code,
repository info, tagged identifiers, the DFG, and the desired (Gold) summary are all included in each few-shot. The target example includes just analysis product, and the LLM is prompted to produce the desired output.

In prior work using ``chain of thought''~\cite{wei2022chain} or ``step by step''~\cite{kojima2022large} reasoning, no such information is given to the model; instead, the prompt simply helps it organize its reasoning about the sample into a sequence of instructions. Here, rather than having the model do its own reasoning, we shape its reasoning externally by using simple program analyses, since
we can get very precise information from very efficient analysis tools. Each few-shot example includes source code, derived information, and conclusion (summary), thus providing  exemplary "chains of thought" for the model to implicitly use when generating the desired target summary. \Cref{pipeline} presents the overall pipeline of our approach that we apply to each sample. The BM25 engine matches input code against a sample pool, \asap processes resulting examples to create a prompt, and the final prompt is sent to the GPT-3.x model via API, yielding a summary as output.
 
Next, we describe how we evaluate this pipeline. 

\vspace{-0.06in}
\subsection{Metrics}

BLEU~\cite{papineni2002bleu} is the most widely-used, similarity-based measure
for code summarization~\cite{roy2021reassessing} and commit log generation~\cite{dey2022evaluating}. 
BLEU counts the \emph{fraction} of $n$-grams (usually for $n \in [1..4]$), that occur in both generated candidates and one or more reference translations;  the geometric mean of these fractions is the BLEU, usually normalized to the range 0-100. At \emph{sentence} granularity, 
BLEU tends to overly penalize candidate translations when few (or none)  of the longer n-grams co-occur,  so
"Sentence BLEU" has been criticized for correlating poorly with human judgment. 
Various smoothing techniques~\cite{chen2014systematic,lin2004orange,gao2013training} have been used, to reduce Sentence BLEU's sensitivity to sparse $n$-gram matches, and better align it with human quality assessment. We report data on two variants: 
BLEU-CN, which uses a kind of Laplacian smoothing~\cite{alon2018code2seq,feng2020codebert,iyer2016summarizing,wang2021codet5,ahmad2021unified,ahmed2022few, lu2021codexglue} 
and BLEU-DC, which uses newer smoothing methods~\cite{hu2018deep,wei2019code}. 
Other proposed metrics such as BERTScore~\cite{zhang2019bertscore,haque2022semantic}, BLEURT~\cite{sellam2020bleurt}, NUBIA~\cite{kane2020nubia},  
are computationally expensive, not widely used and thus
not readily comparable with prior work for benchmarking.  

Given all these options, metrics for code summarization and, independently, 
for commit-log generation~\cite{dey2022evaluating}, have been debated~\cite{gros2020code,roy2021reassessing,haque2022semantic}. 
In this paper, we follow prior work and primarily use BLEU-CN; this
facilitates the comparison of our results with prior work. 
The CodeXGLUE benchmark recommends BLEU-CN, and most newer models~\cite{feng2020codebert, wang2021codet5,ahmed2022few} use this metric. 
We, however, \emph{have not neglected other measures.}
Besides BLEU-CN, and BLEU-DC, 
we also report results using ROUGE-L~\cite{lin2004rouge} and METEOR~\cite{banerjee2005meteor}. 

In all cases, \asap achieves significant overall improvements:
we observe gains greater than 2.0 BLEU for all programming languages except for Go (Table~\ref{tbl:codex}). 
We contend that gains greater than 2.0 BLEU are important for two reasons.
 Roy \etal\cite{roy2021reassessing} provide arguments, grounded on human subject study that \emph{for code summarization} (our central task), that a gain of 2.0 or more BLEU is more likely to correspond with human perception of improvement. 
Second, we argue that {\bf even smaller gains matter} (especially
if repeatable and statistically significant)
since incremental progress on such tasks accumulates, towards strong practical impact, as evidenced by decades-long work in natural language translation. 

In addition to code summarization, we evaluated \asap approach on the code completion task. The standard metrics used for this task are \emph{exact match} (did the completion match 
exactly) and \emph{edit similarity} (how close is the completion to the expected sequence). Here, too, \asap achieves significant overall improvements. 

\subsection{Experimental Setup \& Evaluation Criteria}
Our primary model is OpenAI's code-davinci-002. We use the beta version, via its web service API. To balance computational constraints like rate limits and our desire for robust estimates of performance, we chose to use 1000 samples\footnote{Please see \Cref{sec:sample-size} for
the rationale.}  
per experimental treatment (one treatment for each language, each few-shot selection approach, with \asap, without \asap \etc). 

Our experiments yielded statistically significant, interpretable results in most cases. Each 1000-sample trial still took 5 to 8 hours, varying (presumbly) with 
OpenAI's load factors. We include waiting periods between attempts, following OpenAI's recommendations.  
To obtain well-defined answers from the model, we found it necessary to set the temperature to 0, for all our experiments. The model is designed to allow a window of approximately 4K tokens; this limits the number of few-shot samples. For our experiments, we used 3 shots. \asap defaults to three shots
because related work~\cite{brown2020language,ahmed2022few} has shown, and our own experiments with \asap confirmed,
that more shots did not significantly improve performance. 
However, for up to 2\% of the randomly chosen samples in each experiment, we didn't get good results; either the prompt didn't fit into the model's window, or the model mysteriously generated an empty string. In cases where the prompt as constructed with 3 samples was too long,  we automatically reduce the number of shots. When empty summaries were emitted, we  resolved this by increasing the number of shots. This simple, repeatable, modest-overhead procedure can be incorporated into automated summarization tools.

\begin{table*}[!t]

\resizebox{.90\textwidth}{!}{
\renewcommand{\arraystretch}{1.2}
\begin{tabular}{lccccccccc}
\hline
\multicolumn{1}{c}{Language} & CodeBERT & GraphCodeBERT & \begin{tabular}[c]{@{}c@{}}Polyglot \\ CodeBERT\end{tabular} & \begin{tabular}[c]{@{}c@{}}Polyglot \\ GraphcodeBERT\end{tabular} & CodeT5 & CodeT5+ & \begin{tabular}[c]{@{}c@{}}Few-shot \\ (random)\end{tabular} & \begin{tabular}[c]{@{}c@{}}Few-shot \\ with BM25\end{tabular} & \begin{tabular}[c]{@{}c@{}}Gain (\%) over \\ random few-shot\end{tabular} \\ \hline
Java                         & 18.8     & 18.52         & 20.22                                                        & 19.94                                                             & 19.78 & 19.83 & 19.87                                                        & \textbf{22.87}                                                         & +15.10\%                                                                      \\
Python                       & 17.73    & 17.35         & 18.19                                                        & 18.33                                                             & 19.98 & 18.85  & 20.66                                                        & \textbf{21.78}                                                         & +5.42\%  \\ \hline                                                                    
\end{tabular}
}
\vspace{0.05in}
\caption{Performance of encoder-decoder and few-shot models on Java and Python code summarization, measured using BLEU.}
\label{baseline}
\vspace{-0.2in}
\end{table*}
\section{Results}
\label{result}
We evaluate the benefits of \asap-enhanced prompts, for code summarization, in different settings and using various metrics. We find evidence of overall performance gain, in studies on six languages. However, for other detailed analyses, we focused primarily on Java and Python, because of OpenAI API rate limits.  

\subsection{Encoder-decoders \& Few-shot Learning}
\label{fixed_few_shot}

Our baseline results on CodeSearchNet~\cite{lu2021codexglue}, using IR-based
few-shotting, come first. 
Prior work reports that IR methods can find better samples for few-shot prompting,   
for tasks such as program repair~\cite{nashid2023retrieval} and code generation~\cite{jain2022jigsaw}. 
In Table~\ref{baseline}, we observe that this is also true for code summarization;  
we note improvements of 3.00 (15.10\%) and 1.12 (5.42\%) in BLEU-4 score for Java and Python, respectively, simply by using \emph{BM25} as a few-shot sample selection mechanism. Since \emph{BM25} was already used in prior paper (albeit for other tasks) \cite{nashid2023retrieval}, we consider this \emph{BM25}-based few-shot learning for code
summarization as just a baseline (not a contribution \emph{per se}) of this paper.

\subsection{\asap Prompt Enhancement}
\label{pefl}
We now focus on the central result of our paper: the effect of \asap prompt enhancement. Table~\ref{tbl:codex} shows the  component-wise and overall 
improvements achieved after combining all the prompting components for all six programming languages. BLEU improvements range from 1.84 (8.12\%) to 4.58 (16.27\%). In most cases, we see improvements of over 2.0 BLEU, the required threshold for human perception
noted by 
Roy \etal~\cite{roy2021reassessing}.

We also noticed that all three components (i.e., \emph{Repo}sitory Information., \emph{DFG} Data Flow Graph, \emph{Id}entifiers) help the model achieve better performance in all six languages, as we combined these components individually with \emph{BM25}. However, for Ruby, the best performing combination 
includes just the 
Repo. information. 
In most cases, the Repo. helps a lot, relative to other components. 

To ascertain improvement significance, we used the pairwise one-sided Wilcoxon signed-rank test, finding statistical significance in all cases for our final prompt when compared with vanilla BM25 few-shot learning, even after adjusting for false discovery risk. 

\begin{table*}[h]

\centering

\resizebox{.65\textwidth}{!}{%
\renewcommand{\arraystretch}{1.2}
\begin{tabular}{lccccccc}
\hline
\multicolumn{1}{c}{\multirow{2}{*}{Language}} & \multicolumn{1}{c}{\multirow{2}{*}{BM25}} & \multicolumn{1}{c}{\multirow{2}{*}{BM25+repo}} & \multicolumn{1}{c}{\multirow{2}{*}{BM25+id}} & \multicolumn{1}{c}{\multirow{2}{*}{BM25+DFG}} & \multicolumn{1}{c}{\multirow{2}{*}{\asap}} & \multicolumn{2}{c}{Comparing with BM25}                                        \\ 
\multicolumn{1}{c}{}                          & \multicolumn{1}{c}{}                      & \multicolumn{1}{c}{}                           & \multicolumn{1}{c}{}                         & \multicolumn{1}{c}{}                          & \multicolumn{1}{c}{}                           & \multicolumn{1}{c}{Gain (\%) over BM25} & \multicolumn{1}{c}{p-value}         \\ \hline
Java                                            & 22.87                                      & 25.23                                           & 23.39                                         & 23.13                                          & \textbf{25.41}                                           & +11.11\%                                    & \textless{}0.01                      \\
Python                                          & 21.78                                      & 24.22                                           & 22.54                                         & 21.82                                          & \textbf{24.26}                                           & +11.39\%                                    & \textless{}0.01                      \\
Ruby                                            & 17.21                                      & \textbf{19.67}                                           & 19.19                                         & 17.55                                          & 19.62                                           & +14.00\%                                       & \textless{}0.01                      \\
JavaScript                                      & 23.27                                      & 25.11                                           & 24.21                                         & 24.04                                          & \textbf{25.36}                                           & +8.98\%                                     & \textless{}0.01                      \\
Go                                              & 22.67                                      & 24.41                                           & 23.2                                          & 23.42                                          & \textbf{24.51}                                           & +8.12\%                                    & \textless{}0.01                      \\
PHP                                             & 28.15                                      & 32.07                                           & 29.8                                          & 28.92                                          & \textbf{32.73}                                           & +16.27\%                                    & \textless{}0.01                      \\ \hline
\multicolumn{1}{l}{Overall}                   & \multicolumn{1}{c}{22.66}                 & \multicolumn{1}{c}{25.12}                      & \multicolumn{1}{c}{23.72}                    & \multicolumn{1}{c}{23.15}                     & \multicolumn{1}{c}{\textbf{25.32}}                      & \multicolumn{1}{c}{+11.74\%}               & \multicolumn{1}{c}{\textless{}0.01} \\ \hline
\end{tabular}
}
\vspace{0.05in}
\caption{Performance of prompt enhanced comment generation with code-davinci-002 model, measured using BLEU. p-values are calculated applying one-sided pair-wise Wilcoxon signed-rank test and B-H corrected.}
\label{tbl:codex}
\vspace{-0.2in}
\end{table*}


\begin{table*}[h]

\centering

\resizebox{.90\textwidth}{!}{%
\begin{tabular}{llcccccccc}
\hline
\multirow{2}{*}{Language} & \multirow{2}{*}{Project Name} & \multirow{2}{*}{\#of training sample} & \multirow{2}{*}{\#of test sample} & \multicolumn{3}{c}{Cross-project}                & \multicolumn{3}{c}{Same-project}                 \\
                          &                               &                                       &                                   & BM25  & \asap  & p-value                          & BM25  & \asap  & p-value                          \\ \hline
\multirow{4}{*}{Java}     & wildfly/wildfly               & 14                                    & 100                               & 24.05 & \textbf{24.77} & \multirow{8}{*}{\textless{}0.01} & 17.86 & 18.27 & \multirow{8}{*}{\textless{}0.01} \\
                          & orientechnologies/orientdb    & 10                                    & 100                               & 25.54 & \textbf{27.23} &                                  & 19.43 & 20.24 &                                  \\
                          & ngageoint/geopackage-android  & 11                                    & 100                               & 29.33 & 42.84 &                                  & 45.48 & \textbf{46.21} &                                  \\
                          & RestComm/jain-slee            & 12                                    & 100                               & 17.04 & 19.06 &                                  & 17.99 & \textbf{19.61} &                                  \\
\multirow{4}{*}{Python}   & apache/airflow                & 12                                    & 100                               & 20.39 & 20.37 &                                  & 20.36 & \textbf{20.72} &                                  \\
                          & tensorflow/probability        & 18                                    & 100                               & \textbf{21.36} & 21.18 &                                  & 20.30 & 20.86 &                                  \\
                          & h2oai/h2o-3                   & 14                                    & 100                               & 19.50 & \textbf{20.72} &                                  & 18.75 & 19.81 &                                  \\
                          & chaoss/grimoirelab-perceval   & 14                                    & 100                               & 25.23 & 29.23 &                                  & 32.75 & \textbf{38.23} &                      \\ \hline           
\end{tabular}
}
\vspace{0.05in}
\caption{Performance of prompt enhanced comment generation with code-davinci-002 model on same project data (measured using BLEU) and p-values are calculated applying one-sided pair-wise Wilcoxon signed-rank test after combining the data from all projects.}
\label{tbl:sameProject}
\vspace{-0.2in}
\end{table*}

\begin{table*}[h]

\centering

\resizebox{.80\textwidth}{!}{%
\begin{tabular}{lccccccccc}
\hline
\multicolumn{1}{c}{\multirow{2}{*}{Language}} & \multicolumn{1}{c}{\multirow{2}{*}{\# of Samples}} & \multicolumn{4}{c}{Exact Match (EM)}                                                                                                                                            & \multicolumn{4}{c}{Edit Similarity (ES)}                                                                                                                                        \\ 
\multicolumn{1}{c}{}                          & \multicolumn{1}{c}{}                              & \multicolumn{1}{c}{Zero-shot} & \multicolumn{1}{c}{\begin{tabular}[c]{@{}c@{}}\asap \\ (Zero-shot)\end{tabular}} & \multicolumn{1}{c}{Gain (\%)} & \multicolumn{1}{c}{p-value} & \multicolumn{1}{c}{Zero-shot} & \multicolumn{1}{c}{\begin{tabular}[c]{@{}c@{}}\asap \\ (Zero-shot)\end{tabular}} & \multicolumn{1}{c}{Gain (\%)} & \multicolumn{1}{c}{p-value} \\ \hline
Java                                            & 9292                                               & 20.75                          & \textbf{22.12}                                                                            & +6.6\%                           & <0.01                        & 55.35                          & \textbf{59.66}                                                                            & +7.79\%                          & <0.01                        \\
Python                                          & 6550                                               & 14.05                          & \textbf{14.58}                                                                            & +3.77\%                          & 0.13                         & 49.71                          & \textbf{50.12}                                                                            & +0.82\%                          & <0.01                        \\ \hline
\multicolumn{1}{l}{Overall}                   & \multicolumn{1}{c}{15842}                         & \multicolumn{1}{c}{17.97}     & \multicolumn{1}{c}{\textbf{19.01}}                                                       & \multicolumn{1}{c}{+5.79\%}     & \multicolumn{1}{c}{<0.01}   & \multicolumn{1}{c}{53.01}     & \multicolumn{1}{c}{\textbf{55.72}}                                                       & \multicolumn{1}{c}{+5.11\%}     & \multicolumn{1}{c}{<0.01}   \\ \hline
\end{tabular}
}
\caption{Performance of \asap enhanced prompts with code-davinci-002 model on line completion task. }
\label{tbl:code-completion}
\vspace{-0.3in}
\end{table*}

\subsection{Same Project Code Summarization}
\label{same-project-summary}

We now examine the benefits of \asap in the context
of some earlier work on few-shot selection. 
Prior work has shown that selecting few-shots from
the same projects substantially improves performance~\cite{ahmed2022few}. 
To see if our prompt enhancement idea further helps in project-specific code summarization, we evaluated our approach on the dataset from Ahmed and Devanbu~\cite{ahmed2022few}. Due to rate limits, we reduced the number of test samples to 100 for each of the four Java and Python projects. Since we have too few samples for a per-project test, we combined all the samples to perform the statistical test. 
Note that our total sample size for the statistical test exceeds the number of required samples determined through the analysis mentioned in \Cref{threat}.
When working with the same project, one must split data with care, to avoid leakage from future samples (where desired outputs may already
exist) to past ones. Therefore, we sorted the samples by creation dates in this dataset. After generating the dataset, we applied our approach to evaulate the performance in same project setting.
We also compared our results with a cross-project setup, where we retrieved samples from the complete cross-project training set, similar to the setting used in Section~\ref{pefl}.

Table~\ref{tbl:sameProject} shows the results project-based code summarization. Note that this is a project-specific scenario where data is not available at all. The training data for each project is very limited. We found that, for 4 projects, cross-project few-shot learning yielded the best performance; while, for  4 others, same-project few-shot learning was most effective. 
We note that Ahmed \& Devanbu didn't use IR to select  few-shot samples and consistently achieved better results with same-project few-shot learning~\cite{ahmed2022few}. IR does find relevant examples
in the large samples available for Java \& Python, and we get good results.
We analyzed 16 pairs of average BLEU 
from 8 projects, considering both cross-project and same-project scenarios. Our prompt-enhanced few-shot learning outperformed vanilla BM25 retrieved few-shot learning in 14 cases (87.5\%). This suggests that \asap prompt enhancement is helpful across projects. 
\asap statistically improves performance in both cross-project and same-project settings.

\vspace{-.06in}

\subsection{Is \asap Model-agnostic?}

Our results so far pertain to the code-davinci-002 models. 
We also fed \asap-augmented prompts to the other two models, text-davinci-003 \& gpt-3.5-turbo (chat model). Our findings are in Table~\ref{tbl:turbo}. Our prompt-enhanced few-shot learning approach improved the performance of the gpt-3.5-turbo model by 1.68\% to 9.13\% and test-davinci-003 model by 13.08\% to 18.69\% on 500 samples each from Java, Python, PHP.

Gpt-3.5-turbo does worse than the code-davinci-002 and text-davinci-003  models at code summarization. The Turbo version is verbose and produces comments stylistically different from those written by developers, and also from the few-shot exemplars in the prompt. 
Careful prompt-engineering might improve the turbo model and enable it to generate more natural, brief comments; this is left for future work. This underperformance by the chat model is consistent with the findings by Kocon et al.~\cite{kocon2023chatgpt}.
Text-davinci-003 model showed the maximum performance increase (albeit still outdone by code-davinci-002). Note that text-davinci-003 is a completion model, like code-davinci-002. Our findings suggest that \asap is more effective with completion models than chat models.    
We also conducted pairwise one-sided Wilcoxon signed rank tests, and the statistical significance of our findings (except java with gpt-3.5-turbo) suggests that \asap will apply beyond just the original code-davinci-002 model.  


\begin{table}[t!]
\resizebox{.90\columnwidth}{!}{%
\renewcommand{\arraystretch}{1.2}
\begin{tabular}{llcccc}
\hline
\multicolumn{1}{c}{Language} & \multicolumn{1}{c}{Model} & BM25  & \asap  & Gain   & p-value         \\ \hline
\multirow{3}{*}{Java}        & Code-davinci-002          & 23.90 & \textbf{25.78} & +7.87\%  & \textless{}0.01 \\
                             & Text-davinci-003          & 18.98 & \textbf{22.31} & +17.54\% & \textless{}0.01 \\
                             & Turbo-GPT-3.5             & 16.68 & \textbf{16.96} & +1.68\%  & 0.95            \\ \hline
\multirow{3}{*}{Python}      & Code-davinci-002          & 22.00 & \textbf{24.78} & +12.64\% & \textless{}0.01 \\
                             & Text-davinci-003          & 16.74 & \textbf{18.93} & +13.08\% & \textless{}0.01 \\
                             & Turbo-GPT-3.5             & 15.01 & \textbf{16.38} & +9.13\%  & \textless{}0.01 \\ \hline
\multirow{3}{*}{PHP}         & Code-davinci-002          & 28.42 & \textbf{33.52} & +17.95\% & \textless{}0.01 \\
                             & Text-davinci-003          & 21.67 & \textbf{25.72} & +18.69\% & \textless{}0.01 \\
                             & Turbo-GPT-3.5             & 18.48 & \textbf{19.99} & +8.17\%  & \textless{}0.01 \\ \hline
\end{tabular}
}
\caption{Performance on code summarization, measured using BLEU. p-values are calculated applying one-sided pair-wise Wilcoxon signed-rank test and B-H corrected.}
\label{tbl:turbo}
\vspace{-0.3in}
\end{table}

\vspace{-.06in}
\subsection{\asap for Completion}
Our primary focus so far has been on code-summarization, in a few-shot setting.  Here, we explore if \asap works on another task: \emph{code completion}, in a \emph{zero-shot} setting where no example is shown or presented to the model.  
We assessed the value of including semantic facts for the \emph{line completion} task, where the model generates the next line given the prior line. We uniformly and randomly collected 9292 Java and 6550 Python samples from the CodeSearchNet dataset to conduct our evaluation.
We randomly selected  a line for each sample and tasked the model with generating that line, given \emph{just all the preceding lines}. 
While applying \asap, we append the repository information and other semantic facts (\ie tagged identifiers, DFG)  \emph{before} the preceding lines.
Importantly, when generating tagged identifiers and DFG, we only used partial information
from preceding lines to avoid information leakage from later lines to the target lines.

We used two metrics, Exact Match (EM) and Edit Similarity (ES), in line with the CodeXGLUE benchmark, to measure the model's performance.
We conducted a McNemar test for EM and a pair-wise Wilcoxon sign-rank test to evaluate the model's performance, similar to what we performed for code summarization. \Cref{tbl:code-completion} summarizes our findings. We observe an overall 5.79\% gain in Exact Match (EM) and a 5.11\% gain in Edit Similarity (ES), highlighting the effectiveness of incorporating semantic facts.
For Python, we find statistical significance only for ES improvement, not for EM. 

\subsection{Performance on Other Metrics}
\label{other-met}
In addition to BLEU-CN, we measured performance with 3 other metrics; BLEU-DC, ROUGE-L
and METEOR. 
Our results, in \autoref{metric}, shows average gains with \asap on all three metrics. 
We conducted pairwise one-sided Wilcoxon signed-rank tests and found 
significant performance improvements with BLEU-DC and ROUGE-L for all the languages. However, we did not observe significant differences with METEOR for 4 out of 6 languages, though sample averages do improve with \asap in all 6 comparisons. It's worth noting that we had only 1000 language samples (due to cost) for each language, so it's not unexpected to see some cases where we didn't observe significance. To evaluate the overall impact of \asap, we combined the dataset from all languages for code-davinci-002 model (6000 samples) and performed the same test; we then get statistical significance (p-value < 0.01) for all three metrics, suggesting that \asap does provide value.


\begin{table}[t!]
\centering
\resizebox{.60\columnwidth}{!}{%
\renewcommand{\arraystretch}{1.2}
\begin{tabular}{llc}
\hline
\multicolumn{1}{c}{Language} & Prompt Component & BLEU-4 \\ \hline
\multirow{4}{*}{Java}        & ALL              & \textbf{25.41}  \\
                             & -Repo.           & 23.50   \\
                             & -Id              & 25.27  \\
                             & -DFG             & 24.86  \\ \hline
\multirow{4}{*}{Python}      & ALL              & \textbf{24.26}  \\
                             & -Repo.           & 22.80  \\
                             & -Id              & 23.93  \\
                             & -DFG             & 23.31  \\ \hline 
\end{tabular}
}
\caption{Ablation study.}
\label{ablation}
\vspace{-0.3in}
\end{table}


\begin{table}[t!]
\centering
\resizebox{.98\columnwidth}{!}{%
\renewcommand{\arraystretch}{1.2}
\begin{tabular}{llc}
\hline
\multicolumn{3}{c}{\textbf{Example 1}}                                                                                                                                                                                                                                                                                                                                                                                                                                                                                                                 \\ \hline
\multicolumn{3}{l}{\begin{tabular}[c]{@{}l@{}}def round(input\_a, name: nil)\\   \hspace{.5cm}    check\_allowed\_types(input\_a, TensorStream::Ops::FLOATING\_POINT\_TYPES)\\    \hspace{.5cm}   \_op(:round, input\_a, name: name)\\ end\end{tabular}}                                                                                                                                                                                                                                                                                                                  \\ \hline
\multicolumn{1}{c}{Gold \& model output}                                                                                                                    & \multicolumn{1}{c}{Comment}                                                                                                                                                                                                     & \multicolumn{1}{c}{BLEU}                                                                                                                  \\ \hline
Gold                                                                                                                                                          & \begin{tabular}[c]{@{}l@{}}Rounds the values of a tensor \\ to the nearest integer element - wise\end{tabular}                                                                                                                   & NA                                                                                                                                           \\
BM25                                                                                                                                                          & Round a tensor to the nearest integer                                                                                                                                                                                            & 39                                                                                                                                         \\
\asap                                                                                                                                                          & \begin{tabular}[c]{@{}l@{}}Rounds the values of a tensor to \\ the nearest integer, element-wise.\end{tabular}                                                                                                                   & 74                                                                                                                                         \\ \hline
\multicolumn{3}{c}{\textbf{Example 2}}                                                                                                                                                                                                                                                                                                                                                                                                                                                                                                                 \\ \hline
\multicolumn{3}{l}{\begin{tabular}[c]{@{}l@{}}public static void main(final String{[}{]} args) \{\\ \hspace{.5cm}        loadPropertiesFiles(args);\\      \hspace{.5cm}   final ShutdownSignalBarrier barrier = new ShutdownSignalBarrier();\\        \hspace{.5cm} final MediaDriver.Context ctx = new MediaDriver.Context();\\   \hspace{.5cm}      ctx.terminationHook(barrier::signal);\\    \hspace{.5cm}     try (MediaDriver ignore = MediaDriver.launch(ctx))\\      \hspace{.5cm}   \{\\      \hspace{1cm}       barrier.await();\\       \hspace{1cm}      System.out.println("Shutdown Driver...");\\     \hspace{.5cm}    \}\\     \}\end{tabular}} \\ \hline
\multicolumn{1}{c}{Gold \& model output}                                                                                                                    & \multicolumn{1}{c}{Comment}                                                                                                                                                                                                     & \multicolumn{1}{c}{BLEU}                                                                                                                  \\ \hline
Gold                                                                                                                                                          & \begin{tabular}[c]{@{}l@{}}Start Media Driver as a \\ stand - alone process .\end{tabular}                                                                                                                                       & NA                                                                                                                                           \\
BM25                                                                                                                                                          & \begin{tabular}[c]{@{}l@{}}Main method that starts the\\  CLR Bridge from Java .\end{tabular}                                                                                                                                    & 10                                                                                                                                         \\
\asap                                                                                                                                                          & \begin{tabular}[c]{@{}l@{}}Main method for running Media \\ Driver as a standalone process.\end{tabular}                                                                                                                         & 33 \\ \hline                                                                                                                                       
\end{tabular}
}
\caption{Selected examples, illustrating the effectiveness of \asap enhancement.}
\label{example}
\vspace{-0.20in}
\end{table}

\section{Discussion and Ablation Study}

We now present an ablation study of \asap's design and the particular semantic facts our instantiation of \asap uses 
before comparing \asap's output to our vanilla \emph{BM25} baseline. The primary aim of an ablation study is to 
gauge the contribute of each aspect of a model to the final observed performance
In our study, we removed each semantic component of the enhanced prompt and observed performance. We found that the Repo. component contributes most to the model's performance (Table~\ref{ablation}) both for Java and Python. However, tagged identifier and DFG are also helpful, and the best results were obtained when we combined all three components in the prompt.

\noindent\textbf{Two Illustrative Examples}
When manually examining results, we observed that in several samples, the \asap prompt contained information that was crucial for the summary.
\Cref{example} shows two example results that illustrate this point. 
In the first example, the baseline model failed to generate the term "element-wise". However, our prompted enhanced version capture this important concept, yielding a higher BLEU-4 score of 74.0 compared to the baseline score of 39.0. Similarly, in the second example, the baseline model did not recognize the function as a standalone process, leading to a low BLEU score of 10.0. However, our proposed approach did identify the function as a standalone process, resulting in a higher BLEU score of 33.0.

\noindent\textbf{Does the Model Memorize the Path?}
Of the three semantic facts \asap adds to a prompt, repo. information impacts the model's performance most. This may be due to the fact that Code-Davinci-002 had memorized the specific file paths in our data during pre-training; when we provide the path to the function, perhaps the model just recalls memorized information?
To investigate this question, we change the path representation: we took the repository name and path, split the tokens at "/", and presented the model with a list of tokens. The main idea behind this approach is to diffuse the original representation, and present the model with something not encountered during pre-training. If the model isn't literally memorizing, its performance should not be impacted.
We observed that the differences between both versions were very small. For Java, we gained 0.24 BLEU but, for Python, we lost 0.04 with tokenized paths. 
This suggests a lower risk that the model memorized the path to the function.

\vspace{0.04in}
\noindent\textbf{Is the Identifier Tag Necessary?}
In this paper, we assign roles to the identifiers and tag them as \emph{Function Name, Parameters, Identifier} \etc  in the prompt (See \Cref{component}). Does this explicit tagging actually help performance? To investigate this question, we compare the model's performance when provided with a plain, ``tag-free''  list of identifiers. 
We observed that the tagged identifiers lead to better performance for both Java and Python than a simple tag-free list of identifiers. Our performance metric BLEU increased by 0.41 and 1.22 for Java and Python, respectively, suggesting that  
explicit semantic information does indeed contribute to better model performance.

\begin{table}[t!]
\resizebox{.65\columnwidth}{!}{%
\begin{tabular}{lcccc}
\hline
\multicolumn{1}{c}{\multirow{2}{*}{Language}} & \multicolumn{2}{c}{Prompt Enhanced}                & \multicolumn{2}{c}{Vanilla BM25} \\
\multicolumn{1}{c}{}                          & \#of shots         & BLEU-4                        & \#of shots     & BLEU-4          \\ \hline
\multirow{3}{*}{Java}                         & \multirow{3}{*}{3} & \multirow{3}{*}{\textbf{25.41 }} & 3              & 22.87  \\
                                              &                    &                               & 4              & 23.13     \\
                                              &                    &                               & 5              & 23.20     \\ \hline
\multirow{3}{*}{Python}                       & \multirow{3}{*}{3} & \multirow{3}{*}{\textbf{24.26 }} & 3              & 21.78   \\
                                              &                    &                               & 4              & 21.89     \\
                                              &                    &                               & 5              & 21.74    \\ \hline 
\end{tabular}
}
\caption{Comparing with higher-shots Vanilla BM25.}
\label{tbl:high-shot}
\vspace{-0.39in}
\end{table}

\vspace{0.04in}
\noindent\textbf{What's Better: More Shots or ASAP?} Despite having billions of parameters, LLMs have limited prompt sizes. For example, code-davinci-002 and gpt-3.5-turbo support allow prompt-lengths of just 4k tokens. 
\asap augmentation does consume some of the available prompt length budget! Thus we have two design options: 1) use fewer, \asap-Augmented samples in the prompt or 2) use more few-shot samples \emph{sans} augmentation.
To investigate this, we also tried using 4 and 5 shots (instead of 3) for Java and Python with the code-davinci-002 model. However, \autoref{tbl:high-shot} shows that higher shots using BM25 does not necessarily lead to better performance. With higher shots, there is a chance of introducing unrelated samples, which can hurt the model instead of helping it.

Only for Java did we observe better performance with both 4 and 5 shots compared to our baseline model. However, our proposed technique with just 3-shots still outperforms using BM25 with 5 shots. It's worth noting that the context window of the model is increasing day by day, and the upcoming GPT-4 model will allow us to have up to 32K tokens\footnote{https://platform.openai.com/docs/models/gpt-4}. Therefore, the length limit might not be an issue in the near future. However, our study suggests that Automated Semantic Augmentation will still be a beneficial way to use available prompt length budget; moreover, it stands to reason that constructing more signal-rich, informative prompts will beneficial regardless of length.

\begin{table*}[!t]
\resizebox{.90\textwidth}{!}{
\begin{tabular}{lcccccccccccc}
\hline
\multicolumn{1}{c}{\multirow{2}{*}{Language}} & \multicolumn{4}{c}{BLEU-DC}                                                                                                        & \multicolumn{4}{c}{ROUGE-L}                                                                                                        & \multicolumn{4}{c}{METEOR}                                                                                                         \\ 
\multicolumn{1}{c}{}                          & \multicolumn{1}{c}{BM25}  & \multicolumn{1}{c}{ASAP} & \multicolumn{1}{c}{Gain (\%)} & \multicolumn{1}{c}{p-value}         & \multicolumn{1}{c}{BM25}  & \multicolumn{1}{c}{ASAP} & \multicolumn{1}{c}{Gain (\%)} & \multicolumn{1}{c}{p-value}         & \multicolumn{1}{c}{BM25}  & \multicolumn{1}{c}{ASAP} & \multicolumn{1}{c}{Gain (\%)} & \multicolumn{1}{c}{p-value}         \\ \hline
Java                                            & 14.09                      & \textbf{15.94}                          & +13.13\%                          & \textless{}0.01                      & 36.85                      & \textbf{38.41}                          & +4.23\%                           & \textless{}0.01                      & 35.66                      & \textbf{36.10}                           & +1.23\%                           & 0.32                                 \\
Python                                          & 12.63                      & \textbf{14.49}                          & +14.73\%                          & \textless{}0.01                      & 35.32                      & \textbf{37.74}                          & +6.85\%                           & \textless{}0.01                      & 33.05                      & \textbf{35.63}                          & +7.81\%                           & \textless{}0.01                      \\
Ruby                                            & 9.16                       & \textbf{11.01}                          & +20.2\%                           & \textless{}0.01                      & 28.19                      & \textbf{30.55}                          & +8.37\%                           & \textless{}0.01                      & 27.65                      & \textbf{29.20}                           & +5.61\%                           & 0.03                                 \\
JavaScript                                      & 14.89                      & \textbf{16.71}                          & +12.22\%                          & \textless{}0.01                      & 32.28                      & \textbf{33.88}                          & +4.96\%                           & \textless{}0.01                      & 32.08                      & \textbf{33.02}                          & +2.93\%                           & 0.15                                 \\
Go                                              & 17.10                       & \textbf{18.57}                          & +8.60\%                            & \textless{}0.01                      & 41.04                      & \textbf{42.43}                          & +3.39\%                           & \textless{}0.01                      & 36.78                      & \textbf{37.26}                          & +1.31\%                           & 0.27                                 \\
PHP                                             & 16.97                      & \textbf{20.63}                          & +21.57\%                          & \textless{}0.01                      & 40.48                      & \textbf{44.90}                          & +10.92\%                          & \textless{}0.01                      & 40.14                      & \textbf{43.35}                          & +8.00\%                              & \textless{}0.01                      \\ \hline
\multicolumn{1}{l}{Overall}                   & \multicolumn{1}{c}{14.14} & \multicolumn{1}{c}{\textbf{16.23}}     & \multicolumn{1}{c}{+14.78\%}     & \multicolumn{1}{c}{\textless{}0.01} & \multicolumn{1}{c}{35.69} & \multicolumn{1}{c}{\textbf{37.99}}     & \multicolumn{1}{c}{+6.44\%}      & \multicolumn{1}{c}{\textless{}0.01} & \multicolumn{1}{c}{34.23} & \multicolumn{1}{c}{\textbf{35.76}}     & \multicolumn{1}{c}{+4.47\%}      & \multicolumn{1}{c}{\textless{}0.01} \\ \hline
\end{tabular}
}
\caption{The effectiveness of ASAP in popular code summarization metrics. p-values are calculated applying one-sided pair-wise Wilcoxon signed-rank test and B-H corrected. }
\label{metric}
\vspace{-0.25in}
\end{table*}

\vspace{0.04in}

\begin{table*}[t]
\resizebox{.90\textwidth}{!}{%
\renewcommand{\arraystretch}{1.2}
\begin{tabular}{llll}
\toprule
\multicolumn{1}{c}{Change (BLEU-4)} & \multicolumn{1}{c}{Reference}                                                                                                                                                                                     & \multicolumn{1}{c}{BM25}                                                                                                           & \multicolumn{1}{c}{\asap}                                                                                                                      \\ \midrule
\asap (+47)                        & \begin{tabular}[c]{@{}l@{}}Generates a gradient - noise value from the \\ coordinates of a three - dimensional input value\\  and the integer coordinates of a nearby three\\  - dimensional value .\end{tabular} & \begin{tabular}[c]{@{}l@{}}Computes a value at the point \\ (x, y, z) in a 3D Perlin noise function.\end{tabular}                  & \begin{tabular}[c]{@{}l@{}}Generates a gradient-coherent-noise value from \\ the coordinates of a three-dimensional input value.\end{tabular} \\ \midrule
\asap (+26)                        & Replaces type with mapped type for current path .                                                                                                                                                                 & \begin{tabular}[c]{@{}l@{}}Returns the mapped type for the current \\ path, or the target type if no mapping exists .\end{tabular} & Replaces target type with mapped type if any .                                                                                                \\ \midrule
\asap (+41)                        & convert a dataRootExt to a dataRoot                                                                                                                                                                               & Convert to a DataRoot object .                                                                                                     & Convert a DataRootExt to a DataRoot .                                                                                                         \\ \midrule
\asap (-36)                        & Computes the convex hull of the set of points .                                                                                                                                                                   & Computes the convex hull of the set of points                                                                                      & \begin{tabular}[c]{@{}l@{}}Computes the convex hull of the set of points \\ using Andrew's monotone chain algorithm\end{tabular}              \\ \midrule
\asap (-35)                        & Getter for the Uninstall Actions .                                                                                                                                                                                & Returns the uninstall actions .                                                                                                    & Retrieves the uninstall actions for this deployable unit .                                                                                    \\ \midrule
\asap (-67)                        & Get a column of this matrix .                                                                                                                                                                                     & Get a column of this matrix .                                                                                                      & Return the specified column of this matrix as a column vector .  \\ \bottomrule                                                                            
\end{tabular}                                                 
}
\caption{Examples Showing Strength and Weakness of \asap.}
\label{tbl:asap1}
\vspace{-0.25in}
\end{table*}

\noindent\textbf{What's New in \asap's Output?} 
We add a \emph{pro forma} analysis of a few hand-picked examples, to be consistent with peer-review-required community rituals; however, these analyses are highly anecdotal must be interpreted cautiously. We manually examine several samples to discuss our results in greater detail;  specifically, to answer three questions: to specify 1) the new types of information \asap presents to the LLM and 2) how \asap's summaries differ from those created by existing techniques, and 3) to analyze the errors that \asap introduces. \Cref{tbl:asap1} presents some samples where, for the first three, \asap performed very well compared to our retrieval-based baselines, and for the second three, the baseline performed better than \asap. While we discuss our findings in the context of the provided samples, our observations generalise to other samples.

\vspace{0.05in}
\noindent{\emph{The new types of information \asap presents to the LLM:} }
As discussed in the paper, our primary contribution involves \emph{augmenting} retrieved samples (retrieved using BM25, as per Nashid et al.~\cite{nashid2023retrieval}) with \emph{semantic facts}, resulting in improved performance compared to the base retrieval approach. We add semantic facts related to repository details, identifiers, and data flow graphs to both retrieved samples and input code. As anticipated, the added semantic facts transfer into, and enhance, the model output.

In the first sample, the baseline retrieval-only method fails to capture the term ``gradient'' entirely. However, by incorporating semantic facts, the model successfully recovers the term because it is frequently found in both identifiers and repository names, influencing the model's output.
In the second example, where the goal is to replace rather than simply return, the baseline fails to generate the term ``replace'', despite the clear indication in the function name (``replaceWithMappedTypeForPath''). The data flow between identifiers, provided in the semantic facts, may have helped the model recognize replacement operations.

\vspace{0.05in}
\noindent\emph{How \asap's summaries differ from those created by existing techniques:} 
\noindent Following the above discussion, we observed that \asap is generating more specific information:

\begin{enumerate}
\item It identifies ``gradient'' in sample 1.
\item It suggests changing ``return'' to ``replace'' in another sample (sample 2).
\item It recommends changing "dataroot" to ``datarootext'' in a different sample (sample 3).
\end{enumerate}

These differences were observed across multiple samples when comparing our baseline to \asap. The \asap approach consistently produces more specific information compared to the baseline.

\vspace{0.05in}
\noindent{\emph{Analyze the errors that \asap introduces:}} 
The examined examples suggest that \asap can become too specific, and thus not match the developer-written summary. \asap gets over-specific in the last three examples with ``Andrew's monotone chain algorithm'' and ``deployable unit'', ``column vector''. While these terms are not necessarily incorrect, BLEU-4 drops, because the developer-written summary was more generic.



We also observe quantitatively that \asap induced positive changes in 44\% of the samples. However, the performance also declined for 30\% of the samples, and remained the same on the rest. Compared to our baseline (few-shot learning with BM25-retrieved samples), \asap requires more tokens. The additional token cost, per query (both in terms of monetary cost and performance overhead) is quite modest.
On the other hand, we observe a substantial 12\% overall improvement with \asap using the Codex model. 
\section{Related work}

\subsection{Code Summarization}
Deep learning models have advanced the state-of-the-art in SE tasks such as code summarization. The LSTM model for code summarization was first introduced by Iyer \etal~\cite{iyer2016summarizing}. 
Pre-trained transformer-based~\cite{vaswani2017attention} models such as CodeBERT~\cite{feng2020codebert}, PLBART~\cite{ahmad2021unified}, and CodeT5~\cite{wang2021codet5} have been extensively used on the CodeXGLUE~\cite{lu2021codexglue} code summarization dataset, resulting in significant improvements. However, there is a caveat to using pre-trained language models: although these models perform well, extensive fine-tuning is required, which can be data-hungry \& time-consuming. Additionally, separate models had to be trained for different languages, increasing training costs. To reduce the number of models required, multilingual fine-tuning has been suggested,  to maintain or improve performance while reducing the number of models to one~\cite{ahmed2022multilingual}. However, this approach did not reduce the training time or the need for labeled data. 

LLMs, or large language models, are much larger than these pre-trained models, and are trained on much bigger datasets with a simple training objective --- auto-regressive next-token prediction~\cite{brown2020language}. These models perform surprisingly well on tasks, even without fine-tuning. Just prompting  the model with different questions, while providing a few problem-solution exemplars, is sufficient. Few-shot learning has already been applied to code summarization, and has been found to be beneficial~\cite{ahmed2022few}.

\vspace{-0.06in}
\subsection{Other Datasets}

There are several datasets available for code summarization, in addition to CodeXGLUE~\cite{lu2021codexglue}.
TL-CodeSum~\cite{hu2018summarizing} is a relatively smaller dataset, with around 87K samples, but it does include duplicates, 
which may result in high performance estimates that may not generalize.  
Funcom~\cite{leclair2019neural} is a dedicated dataset with 2.1 million Java functions, but contains duplicates. 
We chose CodeXGLUE (derived from CodeSearchNet) because it is a diverse, multilingual dataset that presents a challenge for models. Even well-trained models like CodeBERT struggle on this benchmark; its performance is particularly poor on languages with fewer training samples.

There has been a lot of work on code summarization, ranging from template matching to few-shot learning. These models use different representations and sources of information to perform well in code summarization. Comparing or discussing all of these models is beyond the scope of this work. 
We note, however, that our numbers represent a new high-point on the widely used CodeXGlue benchmark for code summarization and code-completion; we refer the reader to \url{https://microsoft.github.io/CodeXGLUE/} for a quick look at the leader-board. Our samples are smaller (N=1000), but the estimates, and estimated improvements, are statistically robust (See the sample size discussion in \Cref{sec:sample-size}).

\vspace{-.05in}

\subsection{LLMs in Software Engineering}
Although LLMs are not yet so widely used for code summarization, they are extensively used for code generation~\cite{chen2021evaluating,xu2022systematic,nijkamp2022codegen} and program repair~\cite{jiang2023impact,joshi2022repair,fan2022automated,ahmed2023better}. Models like Codex aim to reduce the burden on developers by automatically generating code or completing lines. Several models such as Polycoder~\cite{xu2022systematic} and Codegen~\cite{nijkamp2022codegen} perform reasonably well, and due to their few-shot learning or prompting, they can be applied to a wide set of problems. However, Code-davinci-002 model generally performs well than those models and allows us to fit our augmented prompts into a bigger window.

Jain et al. proposed supplementing LLM operation with subsequent processing steps based on program analysis and synthesis techniques to improve performance in program snippet generation~\cite{jain2022jigsaw}. Barei{\ss} \etal showed the effectiveness of few-shot learning in code mutation, test oracle generation from natural language documentation, and test case generation tasks~\cite{bareiss2022code}. CODAMOSA~\cite{lemieux2023codamosa}, an LLM-based approach, conducts search-based software testing until its coverage improvements stall, then asks the LLM to provide example test cases for functions that are not covered. By using these examples, CODAMOSA helps redirect search-based software testing to more useful areas of the search space. Jiang et al. evaluated the effectiveness of LLMs for the program repair problem~\cite{jiang2023impact}.

Retrieving and appending a set of training samples has been found to be beneficial for multiple semantic parsing tasks in NLP, even without using LLM~\cite{zemlyanskiy2022generate}. One limitation of this approach is that performance can be constrained by the availability of similar examples. Nashid et al. used a similar approach and gained improved performance in code repair and assertion generation with the help of LLM~\cite{nashid2023retrieval}.
However, none of the above works has attempted to automatically semantically augment the prompt. Note that it is still too early to comment on the full capabilities of these large language models. Our findings so far suggest that augmenting the exemplars
in the prompt with semantic hints helps on the code summarization and code completion tasks; judging the value of \asap in other tasks is left for future work.

\vspace{-0.05in}
\section{Threats \& Limitations}
\label{threat}

A major concern when working with large language models is the potential for test data exposure during training. Sadly, one can't directly check this since the training dataset is not accessible. The model's lower performance with random few-shotting suggests that memorization may not be a major factor. As we incorporate relevant information, the model's performance improves with the amount and quality of information. Had the model already memorized the summaries, it could have scored much higher, even without the benefit of relevant exemplars and semantic augmentation. 

\vspace{0.05in}
\label{sec:sample-size}
\noindent\emph{Sample Size Analysis:}
 We used the observed means and standard deviations to calculate (using G*power~\cite{faul2009statistical,faul2007g}) the required sample sizes, using commonly used values:  $\alpha$ of $0.01$ (desired p-value)  and a $\beta$ of 0.20 (\emph{viz}, a 20\% chance of  NOT discovering an effect, should one exist). For the tests that we used (Wilcoxon Signed-rank test), we found that the needed sample size was always below the sample size we used for our primary studies, \emph{viz.,} 1000. 

\vspace{0.05in}
\noindent\emph{User Study:} We did not conduct a user study for \asap. Thus, the enhancements in metrics presented here may not necessarily translate into improved developer performance. This aspect is left to future work.

Finally: fine-tuning large LMs to use derived semantic facts may improve on our augmented prompting approach, but would be costly.  We will leave its consideration to future research.

\section{Conclusion}
In this paper, we explored the idea of \emph{Automatic Semantic Augmentation of Prompts}, whereby we propose
to enhance few-shot samples in LLM prompts, with tagged facts automatically derived by semantic analysis. This based on an intuition that human developers often scan the code to implicitly extract such facts in the process of code comprehension leading to writing a good summary. While
it is conceivable that LLMs can implicitly infer such facts for themselves, we conjectured that
adding these facts in a formatted style to the exemplars and the target, within the
prompt, will help the LLM organize it's ``chain of thought'' as it seeks to construct a summary. We evaluated this idea a challenging, de-duplicated,  well-curated
CodeSearchNet dataset, on two tasks: code summarization and code completion. 
Our findings indicate that Automated Semantic Augmentation of Prompts is generally helpful. Our estimates suggest it helps surpass state-of-the-art.

\vspace{.2cm}

\noindent{\em Acknowledgements:} We would like to acknowledge National Science Foundation under Grant NSF CCF (SHF-MEDIUM) No. 2107592 and the Intelligence Advanced Research Projects Agency (IARPA) under contract W911NF20C0038 for partial support of this work. Our con- clusions do not necessarily reflect the position or the policy of our sponsors and no official endorsement should be inferred.
\bibliographystyle{ACM-Reference-Format}
\bibliography{sample-base.bib}

\end{document}